\documentclass[a4paper,11pt]{article}
\usepackage{jheppub}
\usepackage{orcidlink}

\title{Transverse Pressure in Clouds of Strings: Letelier--Alencar Revisited and Exact Solutions with Curvature Couplings}

\author[a]{G. Alencar\,\orcidlink{0000-0002-3020-4501}}
\author[a]{and R. N. Costa Filho\,\orcidlink{0000-0002-4654-8691}}

\affiliation[a]{Departamento de F\'isica, Universidade Federal do Cear\'a,\\ Caixa Postal 6030, Campus do Pici, 60455-760 Fortaleza, Cear\'a, Brazil}

\emailAdd{geova@fisica.ufc.br}
\emailAdd{rai@fisica.ufc.br}

\abstract{Clouds of strings describe ensembles of relativistic strings coupled to gravity, and a natural question is which macroscopic stresses such ensembles can sustain. The Letelier--Alencar cloud makes this question concrete: it is an anisotropic source with $p_r=-\rho$ and positive tangential pressure. We show that, for a static spherically symmetric source with $p_r=-\rho$, no ensemble of free Nambu--Goto strings with positive tensions and non-negative weights can produce a tangential pressure: the null energy condition, saturated along the two radial null directions, forces every worldsheet to be radial. The same static source admits instead effective realizations in terms of a string fluid with integrable worldsheets, a constrained scalar triplet and dyonic square-root electrodynamics. We then show that a tangential pressure follows from an explicit worldsheet action in which the strings couple nonminimally to the Ricci scalar, the Ricci tensor and the Riemann tensor. We derive the covariant stress-energy tensor for arbitrary embeddings and show that radial worldsheets remain exact solutions of their equations of motion. The resulting self-gravitating cloud admits an exact two-function family of black holes, expressed in terms of the imaginary error function, whose horizon structure we classify analytically. To first order in the couplings, a metric redefinition removes the Ricci-scalar and Ricci-tensor terms, leaving general relativity with a contact interaction quadratic in the density of strings, and the two-function family is the corresponding geometry in the metric to which the strings couple. A particular relation between the couplings yields a single-function branch. The Riemann coupling drops out of the radial field equations but shifts the Wald entropy, and both branches satisfy the first law at fixed parameters.}

\keywords{Black Holes, Bosonic Strings, Classical Theories of Gravity, Effective Field Theories}

\begin{document}
\maketitle
\flushbottom

\section{Introduction}

Clouds of strings provide a simple coarse-grained description of ensembles of relativistic strings coupled to gravity, and a natural question is which macroscopic stresses such ensembles can sustain. The Letelier--Alencar cloud~\cite{Alencar:2025zyl}, an anisotropic source with $p_r=-\rho$ and positive tangential pressure, makes this question concrete. For free Nambu--Goto strings the answer is restrictive: as we show in this work, for a static spherically symmetric source with $p_r=-\rho$, ensembles with positive tensions and non-negative weights cannot produce a tangential pressure, without assuming that the strings are static, straight or non-intersecting. This obstruction can be overcome when the worldsheets couple nonminimally to the spacetime curvature. The resulting self-gravitating cloud admits an exact two-function family of black holes, with two distinct metric functions in areal-radius gauge, whose horizon structure can be classified analytically. Moreover, the coupling to the Riemann tensor drops out of the radial field equations but still contributes to the Wald entropy.

The cloud of strings, introduced by Letelier~\cite{Letelier:1979ej} and developed as string dust by Stachel~\cite{Stachel:1980zr}, is one of the simplest models of this kind. The same exterior geometry also arises for a global monopole~\cite{Barriola:1989hx}, whose asymptotic stress tensor satisfies $\rho=-p_r=\eta^2/r^2$ and $p_\perp=0$, with $a=8\pi G\eta^2$. The Letelier cloud describes a continuous distribution of Nambu--Goto strings and, in spherical symmetry, leads to the black hole $f(r)=1-a-2M/r$, where the constant $a$ is fixed by the density of strings. The model has been extensively applied to black holes in several contexts~\cite{Ganguly:2014cqa,Panotopoulos:2018law,Singh:2020nwo,Mustafa:2022xod,Ghosh:2014dqa,deMToledo:2018tjq,Toledo:2018hav,Kumar:2025kzt,Ahmed:2025qza,Kumar:2025mpb}, including modified theories of gravity~\cite{MoraisGraca:2017geu}. A fluid of strings, in which a pressure is added by hand, was later proposed by Letelier~\cite{Letelier:1983du} and solved by Soleng~\cite{Soleng:1993yr}.

Recently, together with Landim, we proposed a generalization of the cloud of strings by allowing a \textit{magnetic-like} component $\Sigma_{23}$ of the string bivector, in addition to the \textit{electric-like} component $\Sigma_{01}$ of the original model~\cite{Alencar:2025zyl}. The conservation equations fix the source completely in terms of two constants, $a$ and $c_0$, with $p_r=-\rho$ and a positive tangential pressure. The resulting black hole is given in terms of a hypergeometric function and reduces to the Letelier solution for $c_0\to0$. This solution has since been used and extended in several works~\cite{Silva:2025hwl,Muniz:2025reg,Ahmed:2025nfi,EslamPanah:2026u1,EslamPanah:2026cjx,Silva:2026opt,Zhong:2026kqy,Zhong:2026zrr}, where it is often referred to as the Letelier--Alencar cloud of strings. We adopt this name in what follows. In Ref.~\cite{Alencar:2025zyl}, the constants were interpreted through an analogy with the Dirac--Born--Infeld (DBI) action, with $c_0$ identified with the string length and $a$ with $g_s^2$.

A question left open in Ref.~\cite{Alencar:2025zyl} is the microscopic origin of the magnetic-like component. For a single Nambu--Goto string, the bivector is built from two tangent vectors and is therefore simple, $\Sigma\wedge\Sigma=0$. When $\Sigma_{01}$ and $\Sigma_{23}$ are both nonzero this condition is violated, so the Letelier--Alencar bivector cannot be the tangent bivector of a single worldsheet. It could still be the result of an ensemble of strings, each one with its own simple bivector. This is the question answered by the no-go result stated above.

In this work, we revisit the Letelier--Alencar cloud and ask whether its magnetic-like component admits a microscopic description in terms of strings. We first examine whether ensembles of free Nambu--Goto strings can produce a source with $p_r=-\rho$ and $p_\perp>0$, and clarify the nature of the non-simple bivector of the model. We then look for effective realizations of the source in terms of interacting strings, which relate it to the dual strings of Nielsen and Olesen~\cite{Nielsen:1973qs}, to the coupling of strings to two-form fields~\cite{Kalb:1974yc} and to the DBI description of strings carrying electric flux~\cite{Witten:1995im}. Finally, we investigate whether a nonminimal coupling of the worldsheets to the spacetime curvature can generate a tangential pressure with simple bivectors, and study the resulting black hole solutions.

The paper is organized as follows. In Sec.~\ref{sec:micro} we review the Letelier--Alencar cloud, prove the no-go result and discuss the field-theoretic and interacting-string interpretations. In Sec.~\ref{sec:nonminimal} we introduce the nonminimally coupled strings and derive their source. In Sec.~\ref{sec:bh} we find the exact black hole solutions. Sec.~\ref{sec:conclusions} is devoted to the conclusions. In Appendix~\ref{app:Tmunu} we derive the general stress-energy tensor of the cloud. We use the signature $(-,+,+,+)$. In the explicit solutions we absorb $8\pi G$ into the stress-energy tensor, and in the thermodynamic quantities we set $G=1$.

\section{Microscopic Foundations of the Letelier--Alencar Cloud}\label{sec:micro}

In this section we ask which degrees of freedom can reproduce the Letelier--Alencar source. We first show that free Nambu--Goto strings cannot, and then discuss effective realizations.

\subsection{The Letelier--Alencar Cloud}

The starting point is the Nambu--Goto action~\cite{Nambu:1986ze,Goto:1971ce},
\begin{equation}
S_{NG}=-\mathcal{M}\int\sqrt{-\gamma}\,d\lambda^0d\lambda^1,\qquad
\gamma_{AB}=g_{\mu\nu}\frac{\partial x^\mu}{\partial\lambda^A}\frac{\partial x^\nu}{\partial\lambda^B},
\label{NG}
\end{equation}
and the bivector $\Sigma^{\mu\nu}=\epsilon^{AB}\partial_Ax^\mu\partial_Bx^\nu$, with $\gamma=\frac{1}{2}\Sigma^{\mu\nu}\Sigma_{\mu\nu}$. The stress-energy tensor of the cloud is
\begin{equation}
T^\mu_{\ \nu}=-\rho_p\frac{\Sigma^{\mu\alpha}\Sigma_{\alpha\nu}}{\sqrt{-\gamma}},
\label{TLet}
\end{equation}
where $\rho_p$ is the proper density. The overall sign is fixed so that the energy density is positive in our signature. The bivector obeys $\nabla_\mu(\rho_p\Sigma^{\mu\nu})=0$.

For the metric
\begin{equation}
ds^2=-A(r)dt^2+B(r)dr^2+C^2(r)d\Omega^2,
\label{metric}
\end{equation}
the spherically symmetric solutions of the conservation equation are, for the contravariant components,
\begin{equation}
\rho_p\Sigma^{tr}=\frac{a}{C^2\sqrt{AB}},\qquad
\rho_p\Sigma^{\theta\phi}=\frac{b(r)}{C^4\sin\theta}.
\label{gensol}
\end{equation}
Lowering the indices gives $\rho_p\Sigma_{tr}=-a\sqrt{AB}/C^2$ and $\rho_p\Sigma_{\theta\phi}=b(r)\sin\theta$, which coincide with Eq.~(8) of Ref.~\cite{Alencar:2025zyl}, $\rho_p\Sigma_{01}=a\sqrt{AB}/C^2$ and $\rho_p\Sigma_{23}=b\sin\theta$, up to the sign of the electric-like component. Since the source depends only on $a^2$ and $b^2$, only $|a|$ appears in it. In the orthonormal frame, $\rho_p\Sigma^{\hat t\hat r}=a/C^2$ and $\rho_p\Sigma^{\hat\theta\hat\phi}=b/C^2$. The conservation of $T^\mu_{\ \nu}$ fixes $b=ac_0^2/\sqrt{C^4+c_0^4}$, and the source is
\begin{equation}
\rho=-p_r=\frac{|a|}{C^4}\sqrt{C^4+c_0^4},\qquad
p_\perp=\frac{|a|c_0^4}{C^4\sqrt{C^4+c_0^4}}.
\label{LAsource}
\end{equation}
For $c_0\to0$ this reduces to the Letelier source, $\rho=-p_r=|a|/C^2$ and $p_\perp=0$.

\subsection{A No-Go Result for Nambu--Goto Strings}

The non-simplicity of the bivector does not exclude, by itself, a microscopic origin in terms of many strings. A stronger statement follows from the null energy condition.

At a point of its worldsheet, a Nambu--Goto string contributes
\begin{equation}
T^{\mu\nu}_{\rm string}=\tau\left(u^\mu u^\nu-v^\mu v^\nu\right),\qquad\tau>0,
\label{Tstring}
\end{equation}
where $u^\mu$ and $v^\mu$ are orthonormal timelike and spacelike tangent vectors. Let $k^\mu$ be a null vector and decompose it as $k=k_\parallel+k_\perp$, with $k_\parallel$ tangent and $k_\perp$ normal to the worldsheet. Since $k^2=0$, we get
\begin{equation}
T^{\rm string}_{\mu\nu}k^\mu k^\nu=\tau\left[(u\cdot k)^2-(v\cdot k)^2\right]=\tau\,k_\perp\cdot k_\perp\geq0,
\label{NEC}
\end{equation}
with equality if and only if $k$ is tangent to the worldsheet.

Consider now an ensemble of Nambu--Goto strings, with stress tensor
\begin{equation}
T^{\mu\nu}=\sum_iw_i\tau_i\left(u_i^\mu u_i^\nu-v_i^\mu v_i^\nu\right),
\label{ensemble}
\end{equation}
where $w_i\geq0$ and $\tau_i>0$. In a static spherically symmetric configuration we have $T^\mu_{\ \nu}=\mathrm{diag}(-\rho,p_r,p_\perp,p_\perp)$. For the radial null vectors $k_\pm^{\hat\mu}=(1,\pm1,0,0)$ we get $T_{\mu\nu}k_\pm^\mu k_\pm^\nu=\rho+p_r$. From Eq.~\eqref{NEC}, this is a sum of nonnegative terms. Therefore, the condition $\rho+p_r=0$ requires each term to vanish, which means that both $k_+$ and $k_-$ are tangent to every worldsheet with $w_i>0$. Since two independent null vectors span the tangent plane, every worldsheet is tangent to the $t$--$r$ plane. With this, $u_i$ and $v_i$ have no angular components and we obtain $p_\perp=0$.

This result does not assume that the strings are static, straight or non-intersecting. We conclude that the source \eqref{LAsource} with $c_0\neq0$ cannot be produced by any ensemble of free Nambu--Goto strings with positive tensions. A complementary result follows from the local-equilibrium description of networks of Nambu--Goto strings, in which the fluid admits a foliation by non-interacting effective worldsheets with vanishing transverse pressure~\cite{Schubring:2014kaa}.

It is instructive to write the source \eqref{LAsource}, for $C=r$, as a combination of a radial string cloud and a radial Maxwell-type stress,
\begin{equation}
T^\mu_{\ \nu}=\frac{|a|\,\mathrm{diag}(-1,-1,0,0)}{\sqrt{r^4+c_0^4}}
+\frac{|a|c_0^4\,\mathrm{diag}(-1,-1,1,1)}{r^4\sqrt{r^4+c_0^4}}.
\label{decomp}
\end{equation}
The first term is a cloud of radial strings, finite at the center and approaching $|a|/r^2$ at large $r$. The second term carries the tangential pressure and behaves as a radial flux. It dominates near the center, where it goes as $|a|c_0^2/r^4$, and decays as $r^{-6}$. The two terms are not separately conserved. This suggests that the magnetic-like component describes a field, rather than string worldsheets. We make this precise below.

\subsection{Field-Theoretic Realization}\label{sec:field}

The conservation of the bivector current $\rho_p\Sigma^{\mu\nu}$ is the same condition satisfied by the source of a Kalb--Ramond two-form~\cite{Kalb:1974yc}. It can also be solved identically by writing the current as the dual of a closed two-form. Let $F$ be an Abelian field strength, with $F=dA$ locally, and identify $\rho_p\Sigma^{\mu\nu}={}^\star F^{\mu\nu}$. The conservation equation $\nabla_\mu(\rho_p\Sigma^{\mu\nu})=0$ becomes the Bianchi identity, and the simplicity condition becomes
\begin{equation}
\rho_p^2\,\Sigma_{\mu\nu}\tilde\Sigma^{\mu\nu}\propto F_{\mu\nu}\tilde F^{\mu\nu}\propto\frac{a^2c_0^2}{C^4\sqrt{C^4+c_0^4}}.
\label{FFdual}
\end{equation}
The bivector is simple if and only if $F\wedge F=0$. In this case $F$ has rank two and can be written locally as $F=d\varphi\wedge d\chi$. The surfaces of constant $\varphi$ and $\chi$ foliate spacetime, and their tangent bivector is ${}^\star F$. For the Letelier cloud, $F=a\sin\theta\,d\theta\wedge d\phi$, so that $\varphi=-a\cos\theta$, $\chi=\phi$, and the leaves are the radial worldsheets. When $F\wedge F\neq0$ the two-form has rank four and contains a second term, $E\sqrt{AB}\,dt\wedge dr$, associated with the spheres of constant $t$ and $r$, which are spacelike. In this case ${}^\star F$ is not the tangent bivector of any family of worldsheets.

The dynamics is given by the square-root Lagrangian of Nielsen and Olesen~\cite{Nielsen:1973qs,Nielsen:1973cs},
\begin{equation}
\mathcal{L}=-\sqrt{\frac{1}{2}F_{\mu\nu}F^{\mu\nu}}=-\sqrt{\mathcal B^2-E^2},
\label{sqrtNED}
\end{equation}
where $\mathcal B$ and $E$ are the magnetic and electric fields in the orthonormal frame. For radial fields, the magnetic monopole $\mathcal B=a/C^2$, with $a>0$, plays the role of the electric-like component of the bivector, and the Gauss law gives
\begin{equation}
\partial_r\left(\frac{C^2E}{\sqrt{\mathcal B^2-E^2}}\right)=0\ \rightarrow\ E=\mathcal B\frac{c_0^2}{\sqrt{C^4+c_0^4}},
\end{equation}
which reproduces the profile of $b(r)$. With $W=\sqrt{\mathcal B^2-E^2}$, the stress tensor of Eq.~\eqref{sqrtNED} gives
\begin{equation}
\rho=-p_r=\frac{\mathcal B^2}{W},\qquad p_\perp=\frac{E^2}{W},
\end{equation}
and the fields above reproduce Eq.~\eqref{LAsource} for arbitrary $A(r)$ and $B(r)$. The static source of the Letelier--Alencar cloud therefore coincides with that of the dyonic sector of square-root electrodynamics, with $a$ the magnetic charge and $c_0^2$ the electric charge, and the Letelier cloud is the pure monopole~\cite{Beltracchi:2019eoh}. The stress tensor \eqref{LAsource}, together with the corresponding metric in integral form, was obtained by Mazharimousavi and Halilsoy~\cite{Mazharimousavi:2012zx} in square-root electrodynamics. Their Eqs.~(12)--(14) reproduce this source and geometry for a vanishing cosmological constant, with $\beta=c_0^2$ and $\alpha P/\sqrt2=|a|$, after identifying the mass integration constant. Its relation to the Letelier--Alencar construction was discussed in Ref.~\cite{Alencar:2026dsq}. In the orthonormal frame the quantity \eqref{FFdual} is proportional to $E\mathcal B$, so the bivector is non-simple only when both fields are present.

The algebraic stress-tensor identity holds beyond the static ansatz. For a rank-four current with $W>0$, the current and stress-tensor conservation equations are locally equivalent to the equations of square-root electrodynamics. For $\rho_p>0$, the identification $\rho_p\Sigma={}^\star F$ gives $\rho_p\sqrt{-\gamma}=W$, with $W=\sqrt{F_{\mu\nu}F^{\mu\nu}/2}$, and makes the algebraic stress tensor in Eq.~\eqref{TLet} identical to that of square-root electrodynamics wherever $W>0$. Current conservation gives $dF=0$, and the Bianchi identity then yields
\begin{equation}
\nabla_\mu T^\mu_{\ \nu}=F_{\nu\alpha}\nabla_\mu\!\left(\frac{F^{\mu\alpha}}{W}\right).
\end{equation}
On a patch with $F\wedge F\neq0$, $F_{\nu\alpha}$ is invertible, so stress-tensor conservation is equivalent to the field equation of square-root electrodynamics. In the purely magnetic limit the stress-tensor identity still holds, while the invertibility argument ceases to apply.

This also clarifies the analogy with the DBI action~\cite{Fradkin:1985qd,Leigh:1989jq} used in Ref.~\cite{Alencar:2025zyl}. For a string with a worldsheet gauge field, the Born--Infeld action $-\mathcal M\int\sqrt{-\det(\gamma_{AB}+\mathcal F_{AB})}$ gives, in two dimensions, $\det(\gamma_{AB}+\mathcal F_{AB})=\gamma+\mathcal F_{01}^2$. The Gauss law fixes the conjugate momentum $\Pi$, and the Legendre transform at fixed $\Pi$ gives back the Nambu--Goto action with effective tension $\sqrt{\mathcal M^2+\Pi^2}$, as for strings carrying electric flux~\cite{Witten:1995im,Schwarz:1995dk}. The bivector remains simple and no tangential pressure appears. The same happens for the dyonic sector of four-dimensional Born--Infeld electrodynamics~\cite{Born:1934gh} in the strong-field regime. For radial fields $\det(g_{\mu\nu}+F_{\mu\nu}/\beta)/\det g_{\mu\nu}=(1+\mathcal B^2/\beta^2)(1-E^2/\beta^2)$, and in the strong-field limit the Lagrangian tends to $-\mathcal B\sqrt{\beta^2-E^2}$, the Gauss law gives a constant $E$, and the source is of Letelier type with a rescaled constant and $p_\perp=0$. At leading order in the strong-field expansion, the purely magnetic Born--Infeld Lagrangian reduces to the square-root theory, whose monopole stress tensor reproduces the Letelier cloud with $p_\perp=0$~\cite{Beltracchi:2019eoh}. The full Born--Infeld theory retains a subleading tangential pressure. For a magnetic flux proportional to $n$, however, Born--Infeld electrodynamics has a quadratic energy density at low density, whereas the Letelier--Alencar source has a linear term followed by a cubic correction. The Letelier--Alencar profile is therefore not that of a Born--Infeld monopole.

\subsection{Interacting Strings}\label{sec:interacting}

The no-go result excludes a realization of the Letelier--Alencar source by free Nambu--Goto strings with positive tension. A covariant effective realization follows from the string-fluid formalism of Kopczy\'nski~\cite{Kopczynski:1987wb} and of Schubring and Vanchurin~\cite{Schubring:2014ena}. Let $\Phi^1$ and $\Phi^2$ be two scalar fields that label the worldsheets, and define
\begin{equation}
B=d\Phi^1\wedge d\Phi^2,\qquad
n=\sqrt{\frac{1}{2}B_{\mu\nu}B^{\mu\nu}}.
\label{fluid_flux}
\end{equation}
We take $n>0$ and the surfaces of constant $\Phi^1$ and $\Phi^2$ timelike. The two-form $B$ is simple and closed, and its kernel is the tangent plane of the worldsheets. The normal and tangent projectors are
\begin{equation}
\perp^{\mu\nu}=\frac{B^{\mu\alpha}B^\nu_{\ \alpha}}{n^2},\qquad
P^{\mu\nu}=g^{\mu\nu}-\perp^{\mu\nu},
\end{equation}
and $n$ is the density of strings, called the string flux in Ref.~\cite{Schubring:2014ena}.

For an energy density $\mathcal E(n)=n\,\mu(n)$, where $\mu(n)$ is an effective tension that depends on the density of neighboring strings, we consider the action
\begin{equation}
S_{\rm eff}=-\frac{1}{8\pi G}\int d^4x\sqrt{-g}\,\mathcal E(n),
\label{fluid_action}
\end{equation}
where the factor $8\pi G$ is absorbed into $\mathcal E$, as in the explicit solutions. At fixed $\Phi^A$ we have $\delta n=-\frac{1}{2}n\perp^{\mu\nu}\delta g_{\mu\nu}$, and the variation with respect to the metric gives
\begin{equation}
8\pi G\,T^{\mu\nu}=-\mathcal E P^{\mu\nu}+\left(n\mathcal E'-\mathcal E\right)\perp^{\mu\nu}.
\label{fluid_tensor}
\end{equation}
For radial worldsheets this gives
\begin{equation}
\rho=-p_r=\mathcal E(n),\qquad p_\perp=n\mathcal E'(n)-\mathcal E=n^2\mu'(n).
\label{GibbsDuhem}
\end{equation}
This is the Gibbs--Duhem relation of a two-dimensional gas in the plane transverse to the strings. A constant $\mu$ gives the Letelier cloud, and a tension that depends on the density, that is, an interaction between strings, generates the tangential pressure.

Locally on the sphere, we choose $\Phi^1=-\cos\theta$ and $\Phi^2=\phi$. Then $B=\sin\theta\,d\theta\wedge d\phi$ and $n=1/r^2$. The equations obtained by varying $\Phi^A$ in Eq.~\eqref{fluid_action},
\begin{equation}
\nabla_\mu\left(\frac{\mathcal E'}{n}B^{\mu\nu}\partial_\nu\Phi^A\right)=0,\qquad A=1,2,
\end{equation}
are satisfied for an arbitrary static spherically symmetric metric, and the conservation equation $\rho'+2(\rho+p_\perp)/r=0$ follows identically from Eq.~\eqref{GibbsDuhem}. Conversely, any conserved static source with $p_r=-\rho$ is of this form, consistent with generalized string-fluid equations of state~\cite{Santos:2025fluid}. In the radial magnetic sector, the action and the stress tensor coincide with those of a nonlinear electrodynamics whose field strength is identified with $B$~\cite{Bronnikov:2000vy}. For the Letelier cloud, $\mathcal E\propto n$, this is the purely magnetic square-root theory of Sec.~\ref{sec:field}~\cite{Beltracchi:2019eoh}. The fluid formulation restricts $B$ to the form \eqref{fluid_flux}, which must be taken into account when comparing perturbations.

In particular, the choice
\begin{equation}
\mathcal E(n)=|a|\,n\sqrt{1+c_0^4n^2}
\label{fluid_LA_energy}
\end{equation}
reproduces the source \eqref{LAsource}. This gives a covariant effective action for the Letelier--Alencar source with integrable worldsheets and a simple two-form. The non-simple bivector of the original construction, or the dyonic field of Sec.~\ref{sec:field}, and the simple flux of this fluid are distinct realizations of the same static source, and they may differ for perturbations. As for the global monopole~\cite{Barriola:1989hx}, the static source alone does not determine its microscopic origin. The energy function \eqref{fluid_LA_energy} is an effective choice, and its microscopic origin is not determined by this construction.

Expanding $\mathcal E=\sum_kb_kn^k$, the Einstein equations give
\begin{equation}
f(r)=1-b_1-\frac{2M}{r}+\frac{b_2}{r^2}+\frac{b_3}{3r^4}+\frac{b_4}{5r^6}+\dots
\label{virial}
\end{equation}
The linear term describes free strings and gives the solid angle deficit. The quadratic term gives a Reissner--Nordstr\"om-like term, and the cubic term gives a $1/r^4$ term. This expansion classifies the dependence of the tension on the density, and a microscopic model is needed to identify the corresponding interactions between strings. The fluid of Soleng~\cite{Soleng:1993yr}, with $p_\perp\propto\rho$, corresponds to $\mathcal E\propto n^{1+\alpha}$. For the Letelier--Alencar cloud,
\begin{equation}
\mathcal E(n)=|a|n+\frac{|a|c_0^4}{2}n^3+\dots,
\end{equation}
which has no quadratic term: the first correction is cubic in the density and is resummed by the square root.

The energy conditions acquire a simple meaning in this language. Since $\rho+p_r=0$ and $\rho+p_\perp=n\mathcal E'(n)$, the null energy condition requires $\mathcal E'(n)=\mu+n\mu'\geq0$, and the weak energy condition requires in addition $\mu\geq0$. The strong energy condition requires also $\rho+p_r+2p_\perp=2n^2\mu'\geq0$, that is, a repulsive interaction. The dominant energy condition, $|p_\perp|\leq\rho$, requires $|n\mu'|\leq\mu$, that is, the tension can grow at most linearly with the density. Maxwell-type stresses saturate this bound, and the Letelier--Alencar cloud approaches it near the center.

A further effective realization is provided by a constrained scalar triplet $q^a$ with a noncanonical kinetic function, within the class of gravitating nonlinear sigma models studied, for example, in Ref.~\cite{Prasetyo:2017rij}. With internal Euclidean metric $\delta_{ab}$, the action is
\begin{equation}
S_\sigma=-\frac{1}{8\pi G}\int d^4x\sqrt{-g}\left[\mathcal K(Y)+\frac{\ell}{2}\left(q^aq^a-1\right)\right],
\label{sigma_action}
\end{equation}
with $Y=\frac{1}{2}\nabla_\mu q^a\nabla^\mu q^a$ and $\ell$ a Lagrange multiplier. The scalar equations are
\begin{equation}
\nabla_\mu\left(\mathcal K'\nabla^\mu q^a\right)=\ell q^a,\qquad q^aq^a=1.
\end{equation}
For the spherical configuration $q^a=(\sin\theta\cos\phi,\sin\theta\sin\phi,\cos\theta)$ we have $Y=1/r^2$, and the scalar equations are satisfied with $\ell=-2Y\mathcal K'$ for any static spherically symmetric metric written in terms of the areal radius. On the constraint surface,
\begin{equation}
8\pi G\,T_{\mu\nu}=\mathcal K'\nabla_\mu q^a\nabla_\nu q^a-\mathcal K g_{\mu\nu},
\end{equation}
so that $\rho=-p_r=\mathcal K(Y)$ and $p_\perp=Y\mathcal K'-\mathcal K$. For $\mathcal K(Y)=|a|Y$, the constrained model gives $\rho=-p_r=|a|/r^2$ and $p_\perp=0$, reproducing the Letelier geometry. This is the nonlinear-sigma-model realization of the global-monopole exterior, related to the geometry of Barriola and Vilenkin~\cite{Barriola:1989hx,Tan:2017egu}. The choice $\mathcal K(Y)=|a|Y\sqrt{1+c_0^4Y^2}$ reproduces the Letelier--Alencar source exactly. This provides another minimally coupled effective realization of the same static source, without determining its microscopic origin.

The density-dependent tension above is an effective description, which follows from an action for the coarse-grained fluid but not for the individual worldsheets. We next derive transverse stresses from curvature couplings in a worldsheet action.

\section{Nonminimal Curvature Couplings of String Worldsheets}\label{sec:nonminimal}

\subsection{Action}

We now ask how a tangential pressure can be generated from an explicit worldsheet action. The no-go result relies on the algebraic form \eqref{Tstring} of the Nambu--Goto stress tensor. A natural way to escape it, keeping simple bivectors, is to allow the string to couple to the spacetime curvature through its worldsheet. Corrections to the Nambu--Goto action involving the geometry of the worldsheet appear in the effective description of cosmic strings~\cite{Gregory:1988qv,Anderson:1997ip} and in rigid strings~\cite{Polyakov:1986cs,Kleinert:1986bk}. Let $P^{\mu\nu}=\gamma^{AB}\partial_Ax^\mu\partial_Bx^\nu$ be the projector on the worldsheet. Using only the metric and the projector, there are three scalars linear in the curvature~\cite{Fonda:2016ine,Armas:2017pvj}, namely $R$,
\begin{equation}
\mathcal I_2=R_{\mu\nu}P^{\mu\nu},\qquad
\mathcal I_3=R_{\mu\nu\rho\sigma}P^{\mu\rho}P^{\nu\sigma}.
\end{equation}
We do not consider terms with derivatives of the curvature or with odd parity. We then consider the action
\begin{align}
S=&\frac{1}{16\pi G}\int d^4x\sqrt{-g}\,R\nonumber\\
&-\mu\sum_i\int d^2\xi\sqrt{-\gamma_i}\left[1+\lambda_1R+\lambda_2\mathcal I_2+\lambda_3\mathcal I_3\right],
\label{action}
\end{align}
where the sum runs over the strings of the cloud. For $\lambda_1=\lambda_2=\lambda_3=0$ we recover a cloud of Nambu--Goto strings. The couplings have dimensions of length squared and parametrize nonminimal couplings between the worldsheets and the spacetime curvature. If the action is regarded as an effective description, it is reliable when $|\lambda_i\mathcal R|\ll1$ for the relevant curvature invariants $\mathcal R$. The exact solutions found below remain solutions of the action, but their extrapolation to the singular center lies outside this regime. To first order, the $\lambda_1$ and $\lambda_2$ couplings are equivalent to a contact interaction between the strings, as shown in Sec.~\ref{sec:redef}. The self-gravitating solutions below are constructed for a continuum cloud with a smooth angular density. Evaluating the curvature sourced by an individual idealized thin string on its own worldsheet would require a separate regularization prescription. The first two couplings are worldsheet versions of the nonminimal curvature-matter couplings studied in modified gravity~\cite{Bertolami:2007gv,Harko:2010mv,Haghani:2013oma,Odintsov:2013iba}. For a Nambu--Goto cloud the matter Lagrangian density is $-\rho$, defined in Eq.~\eqref{densities} below, and $T^{\mu\nu}=-\rho P^{\mu\nu}$, so that the $\lambda_1$ and $\lambda_2$ terms correspond to couplings of the form $RL_m$ and $R_{\mu\nu}T^{\mu\nu}$. Here, however, the couplings are defined on each worldsheet, and the variation of the projector follows from the embedding, with no additional assumption about the variation of the matter fields. The $\lambda_3$ coupling has no analogue in these theories. In these theories the coupling generically produces an extra force on the matter~\cite{Bertolami:2007gv}. Here the corresponding effect is contained in the equations of motion of the strings, discussed in Sec.~\ref{sec:eom}. In what follows we derive the contribution of each coupling to the source and then consider their combined effect on a radial cloud.

\subsection{Stress-Energy Tensor}

Defining the densities
\begin{equation}
\rho=\frac{\mu}{\sqrt{-g}}\sum_i\int d^2\xi\sqrt{-\gamma_i}\,\delta^{(4)}(x-X_i),\quad
J^{\mu\nu}=\rho P^{\mu\nu},
\label{densities}
\end{equation}
we vary the action with respect to $g_{\mu\nu}$, keeping the embeddings fixed. Besides the variation of the curvature, we must take into account the variation of the projector, $\delta P^{\mu\nu}=-P^{\mu\alpha}P^{\nu\beta}\delta g_{\alpha\beta}$, and of the induced volume element, $\delta\sqrt{-\gamma}=\frac{1}{2}\sqrt{-\gamma}P^{\alpha\beta}\delta g_{\alpha\beta}$. For the first two couplings we obtain
\begin{align}
T^{\mu\nu}=&-\rho P^{\mu\nu}+\lambda_1\Big[-\rho RP^{\mu\nu}+2\rho R^{\mu\nu}\nonumber\\
&+2(g^{\mu\nu}\Box-\nabla^\mu\nabla^\nu)\rho\Big]
+\lambda_2\Big[2\rho R_{\alpha\beta}P^{\alpha(\mu}P^{\nu)\beta}\nonumber\\
&-\rho\,\mathcal I_2P^{\mu\nu}-2\mathcal D^{\mu\nu}[J]\Big],
\label{Tcov}
\end{align}
with
\begin{align}
\mathcal D^{\mu\nu}[J]=&\frac{1}{2}\Big[\nabla_\alpha\nabla^\mu J^{\alpha\nu}+\nabla_\alpha\nabla^\nu J^{\alpha\mu}\nonumber\\
&-\Box J^{\mu\nu}-g^{\mu\nu}\nabla_\alpha\nabla_\beta J^{\alpha\beta}\Big].
\end{align}
The operator $\mathcal D^{\mu\nu}[J]$ follows from the Palatini identity for $\delta R_{\alpha\beta}$ after two integrations by parts, and the order of the covariant derivatives must be kept.

The third coupling is better understood through the contracted Gauss identity~\cite{Capovilla:1994bs,Carter:1997pb},
\begin{equation}
\mathcal I_3={}^{(2)}\!R-K^\mu K_\mu+K^\mu_{AB}K_\mu^{AB},
\label{Gauss}
\end{equation}
where ${}^{(2)}\!R$ is the scalar curvature of the induced metric, $K^\mu_{AB}$ is the extrinsic curvature of the worldsheet and $K^\mu=\gamma^{AB}K^\mu_{AB}$. In two dimensions the Einstein--Hilbert density $\sqrt{-\gamma}\,{}^{(2)}\!R$ is locally a total derivative. Its metric variation contributes only boundary terms, and its Euler--Lagrange derivative vanishes identically. Therefore, up to boundary terms, the $\lambda_3$ coupling is equivalent to the extrinsic curvature term $\lambda_3(K^\mu_{AB}K_\mu^{AB}-K^\mu K_\mu)$. Since this term is quadratic in $K^\mu_{AB}$, its contribution to $T^{\mu\nu}$ vanishes for a cloud of totally geodesic worldsheets, $K^\mu_{AB}=0$. For deformed worldsheets it gives extrinsic-curvature contributions in curved backgrounds, and it is not redundant. In flat spacetime $\mathcal I_3=0$, and the extrinsic combination in Eq.~\eqref{Gauss} reduces to the topological term. A general variational principle for surface actions depending on the extrinsic curvature and on the background curvature, together with the associated spacetime stress tensor, was developed in Ref.~\cite{Armas:2017pvj}, based on the formalism of Carter~\cite{Carter:1997pb}. The explicit derivation of Eq.~\eqref{Tcov} and the general expression for the $\lambda_3$ contribution, in the conventions used here, are given in Appendix~\ref{app:Tmunu}.

\subsection{Equations of Motion of the Strings}\label{sec:eom}

The source \eqref{Tcov} must be supplemented by the equations of motion of the strings, obtained by varying the embeddings $X^\mu_i$ in the action \eqref{action}~\cite{Capovilla:1994bs,Carter:1997pb}. These equations are not automatically satisfied, since the curvature terms act as a position-dependent tension. The diffeomorphism invariance of the matter action implies that $\nabla_\mu T^{\mu\nu}=0$ whenever the equations of motion of the strings hold. Therefore, a consistent solution requires that the embeddings solve their own equations, and not only that the Einstein equations are satisfied.

From Eq.~\eqref{Gauss}, the variation of the $\lambda_3$ term with respect to the embedding is also at least linear in $K^\mu_{AB}$, up to boundary terms. Therefore, totally geodesic worldsheets are not affected by this coupling. This is no longer true for nonlinear functions of $\mathcal I_3$. For example, $\sqrt{-\gamma}\,({}^{(2)}\!R)^2$ is not a total derivative. We comment on this case in Sec.~\ref{sec:conclusions}.

\subsection{Radial Cloud}

We now consider radial strings uniformly distributed over the angles, $\rho=q/(4\pi r^2)$, and the metric
\begin{equation}
ds^2=-N^2(r)F(r)dt^2+\frac{dr^2}{F(r)}+r^2d\Omega^2.
\label{metricNF}
\end{equation}
The radial worldsheets, $X^\mu=(t,r,\theta_0,\phi_0)$, are totally geodesic for arbitrary $N$ and $F$, since they are the fixed points of the rotations around the axis $(\theta_0,\phi_0)$. In this case $\mathcal I_3={}^{(2)}\!R$. We have checked that the equations of motion of the strings, obtained by varying $X^\theta$ and $X^\phi$ in the action \eqref{action} with the three couplings, are identically satisfied for arbitrary $N$ and $F$. The variations of $X^t$ and $X^r$ are tangent to the worldsheet and give no equation. Therefore, the radial strings remain exact solutions with the curvature couplings, without using the Einstein equations.

Defining $a=2Gq$ and $s=a(3\lambda_1+\lambda_2)$, the nonzero components of $8\pi G\,T^\mu_{\ \nu}$ are
\begin{align}
T^t_{\ t}=&-\frac{a}{r^2}+\frac{2(sF-a\lambda_1)}{r^4},\nonumber\\
T^r_{\ r}=&-\frac{a}{r^2}-\frac{2(sF+a\lambda_1)}{r^4},\nonumber\\
T^\theta_{\ \theta}=&\,T^\phi_{\ \phi}=\frac{2a\lambda_1}{r^4}-\frac{2s}{r^4}\left(rF'+rF\frac{N'}{N}-F\right).
\label{Tcomp}
\end{align}
As expected, $\lambda_3$ does not appear. We have checked that this source satisfies $\nabla_\mu T^{\mu\nu}=0$ identically, for arbitrary $N$ and $F$, in agreement with the fact that the radial strings solve their equations of motion. For $\lambda_1=\lambda_2=0$ we recover the Letelier source. The Einstein equations reduce to
\begin{equation}
\frac{N'}{N}=-\frac{2s}{r^3},\qquad
rF'+\left(1-\frac{2s}{r^2}\right)F=1-a-\frac{2a\lambda_1}{r^2},
\label{EinsteinNF}
\end{equation}
and the angular equation follows from the conservation of $T^{\mu\nu}$. Note that $\rho+p_r=-4sF/r^4$, so that $s=0$ is precisely the condition for a metric with a single function.

\subsection{Metric Redefinition}\label{sec:redef}

A natural question is which part of the couplings can be removed by a redefinition of the metric. The couplings $\lambda_1$ and $\lambda_2$ multiply the Ricci scalar and the Ricci tensor, which are fixed by the leading-order field equations in terms of the matter content, and operators of this type can be traded for matter self-interactions by a field redefinition~\cite{Georgi:1991ch,Arzt:1993gz}. This is the worldsheet version of the statement that, for a point particle, the worldline couplings to $R$ and $R_{\mu\nu}u^\mu u^\nu$ can be removed by a redefinition of the metric~\cite{Goldberger:2004jt}. Explicitly, to first order in $\lambda_1$ and $\lambda_2$, the first two couplings in Eq.~\eqref{action} can be written as $-\int d^4x\sqrt{-g}\,R_{\mu\nu}X^{\mu\nu}$, with $X^{\mu\nu}=\rho(\lambda_1g^{\mu\nu}+\lambda_2P^{\mu\nu})$. Under $g_{\mu\nu}=\hat g_{\mu\nu}+h_{\mu\nu}$, the Einstein--Hilbert term changes by $-(16\pi G)^{-1}\int d^4x\sqrt{-\hat g}\,\hat G^{\mu\nu}h_{\mu\nu}$, and the Nambu--Goto term by $-\frac12\int d^4x\sqrt{-\hat g}\,\hat\rho\hat P^{\mu\nu}h_{\mu\nu}$, as follows from Eq.~\eqref{deltarho}. Choosing
\begin{equation}
h_{\mu\nu}=16\pi G\rho\left[\lambda_1g_{\mu\nu}+\lambda_2\left(g_{\mu\nu}-P_{\mu\nu}\right)\right]
\label{redef}
\end{equation}
cancels the $\lambda_1$ and $\lambda_2$ terms, and the action becomes
\begin{equation}
S=\frac{1}{16\pi G}\int d^4x\sqrt{-\hat g}\,\hat R-\int d^4x\sqrt{-\hat g}\,\hat\rho\left[1+16\pi G\lambda_1\hat\rho+\lambda_3\hat{\mathcal I}_3\right]+\mathcal O(\lambda^2).
\label{Sredef}
\end{equation}
Since $P^{\mu\nu}(g_{\mu\nu}-P_{\mu\nu})=0$, the $\lambda_2$ coupling disappears completely at this order, while $\lambda_1$ leaves a contact interaction quadratic in the density of strings. For thin strings it acts only where worldsheets overlap, and for the continuum cloud it is a tension linearly dependent on the density, as in Sec.~\ref{sec:interacting}. It contributes $16\pi G\lambda_1\hat\rho^2$ to the energy density and to the tangential pressure, and the opposite amount to the radial pressure, where $\hat\rho$ is the density of strings of Eq.~\eqref{densities}. This is the quadratic term of Eq.~\eqref{virial}. The $\lambda_3$ coupling involves the Riemann tensor and is not removed in this way, but it does not contribute for radial worldsheets.

For the radial cloud, $8\pi G\hat\rho=a/\hat r^2$, and the metric $\hat g$ is the solution of general relativity with the source above,
\begin{equation}
\hat f(\hat r)=1-a-\frac{2M}{\hat r}+\frac{2a^2\lambda_1}{\hat r^2}.
\end{equation}
Equation~\eqref{redef} rescales $\hat g_{tt}$ and $\hat g_{rr}$ by $1+2a\lambda_1/\hat r^2$ and the angular part by $1+2a(\lambda_1+\lambda_2)/\hat r^2$, so that the areal radius of $g$ is $r^2=\hat r^2+2a(\lambda_1+\lambda_2)$. Expanding to first order in the couplings, we find that $g$ reproduces both $N=e^{s/r^2}$ and the function $F$ of Sec.~\ref{sec:bh}, for all $r$ and with the same $M$. To first order, therefore, the curvature couplings describe general relativity with a contact interaction between the strings, written in the metric to which the strings couple. The difference between the two metric functions reflects the difference between $g$ and $\hat g$. It is physical when other matter couples minimally to $g$, since in the metric $\hat g$ this matter acquires couplings to the density of strings through $h_{\mu\nu}$. Beyond first order, the solutions of Sec.~\ref{sec:bh} are exact solutions of the action \eqref{action} as it stands.

\section{Black Hole Solutions}\label{sec:bh}

\subsection{The Single-Function Branch}

For $\lambda_2=-3\lambda_1$ we have $s=0$ and $N=1$. In this case the dependence on the metric drops out of Eq.~\eqref{Tcomp}, and the source becomes
\begin{equation}
\rho=-p_r=\frac{a}{r^2}+\frac{Q^2}{r^4},\qquad p_\perp=\frac{Q^2}{r^4},\qquad Q^2=2a\lambda_1.
\label{RNsource}
\end{equation}
The solution is
\begin{equation}
f(r)=1-a-\frac{2M}{r}+\frac{2a\lambda_1}{r^2}.
\label{RNlike}
\end{equation}
This source has $p_r=-\rho$ and $p_\perp>0$ for $\lambda_1>0$, as in the Letelier--Alencar cloud, but now with a simple bivector and a source derived from the action. The no-go result does not apply because the strings are no longer described by the Nambu--Goto action alone. In the language of Sec.~\ref{sec:interacting}, Eq.~\eqref{RNlike} corresponds to $\mathcal E=an+Q^2n^2$, that is, to a quadratic correction in the string density in the equivalent fluid description. This is the contact interaction of Eq.~\eqref{Sredef}, whose coefficient $2a^2\lambda_1$ in $\hat f$ becomes $2a\lambda_1$ after the redefinition \eqref{redef}. For the Letelier--Alencar cloud, instead, the first correction is cubic. The nonminimal couplings therefore generate a tangential pressure from a worldsheet action, but not the equation of state of the Letelier--Alencar cloud.

For $a>0$ and $\lambda_1>0$ all the energy conditions are satisfied. For $0<a<1$, $M>0$ and $M^2\geq2a(1-a)\lambda_1$, the horizons are located at
\begin{equation}
r_\pm=\frac{M\pm\sqrt{M^2-2a(1-a)\lambda_1}}{1-a},
\end{equation}
with an extremal black hole when the equality holds,
and the Hawking temperature is
\begin{equation}
T_H=\frac{(1-a)r_+^2-2a\lambda_1}{4\pi r_+^3}.
\end{equation}
Since the matter action depends on the curvature, the entropy must be computed with the Wald formula~\cite{Wald:1993nt,Iyer:1994ys}. Since the horizon binormal lies in the plane of the radial worldsheets, each of the three couplings contributes $-4\pi\lambda_i\rho$ per unit area of the horizon. We find $S=\pi\left[r_+^2-2a(\lambda_1+\lambda_2+\lambda_3)\right]$. With $\lambda_2=-3\lambda_1$ this gives $S=\pi(r_+^2+4a\lambda_1-2a\lambda_3)$, which differs from the area law by a constant. Note that $\lambda_3$ contributes to the entropy, although it does not contribute to the field equations. This is analogous to the Gauss--Bonnet term in four dimensions. Since the correction is a constant, the first law $dM=T_HdS$ is satisfied for fixed $a$ and $\lambda_i$. In this ensemble the first law does not test the constant shift, which follows from the Wald formula itself. The contributions of $\lambda_1$ and $\lambda_2$ coincide with the change of the horizon area under the redefinition of Sec.~\ref{sec:redef}, $\pi\hat r_+^2=\pi[r_+^2-2a(\lambda_1+\lambda_2)]$, as expected, since the entropy can be evaluated in the theory related by the field redefinition~\cite{Jacobson:1993vj}, while the contribution of $\lambda_3$ is not removed.

Due to the solid angle deficit, the spacetime is not asymptotically flat, and some care is needed with the meaning of $M$. With $\tilde t=\sqrt{1-a}\,t$ and $\tilde r=r/\sqrt{1-a}$, the metric becomes
\begin{equation}
ds^2=-\tilde f d\tilde t^2+\frac{d\tilde r^2}{\tilde f}+(1-a)\tilde r^2d\Omega^2,
\end{equation}
with $\tilde f=1-2M/[(1-a)^{3/2}\tilde r]+2a\lambda_1/[(1-a)^2\tilde r^2]$. The coefficient of $1/\tilde r$ is therefore $-2M/(1-a)^{3/2}$, and $M/(1-a)^{3/2}$ is the mass parameter in the normalized coordinates. In this work $M$ denotes the parameter in the metric, with the time $t$. A different normalization of time rescales the energy and $T_H$ by the same factor and does not affect the first law. If the Brown--York energy~\cite{Brown:1992br} is computed with the conical background $f_0=1-a$ as reference, we find $E=M/\sqrt{1-a}$ for the normalized time $\tilde t$, and $E=M$ for the Killing vector $\partial_t$. A complete Hamiltonian derivation of the charges, including the curvature couplings, is beyond the scope of this work.

We stress that, in the static spherically symmetric sector, this source coincides with that of a Letelier cloud plus a magnetic monopole of Maxwell theory with charge $Q$. The geometry \eqref{RNlike} is therefore the known charged black hole surrounded by a cloud of strings~\cite{Toledo:2018hav}, which is also the magnetic sector of the square-root theory combined with Maxwell electrodynamics~\cite{Guendelman:2024kfo}. The difference is the origin of the $1/r^2$ term, which here comes from the strings themselves and not from a gauge field. The two descriptions are expected to differ for perturbations that break the spherical symmetry.

\subsection{The General Branch}

For $s\neq0$, the coupled Einstein--string equations admit the two-function family. Integrating Eq.~\eqref{EinsteinNF} gives $N=e^{s/r^2}$, normalized by $N\to1$ at infinity, and
\begin{equation}
F(r)=1-a+\frac{e^{-s/r^2}}{r}\left[-2M+\mathcal C\,\mathrm{erfi}\!\left(\frac{\sqrt{s}}{r}\right)\right],
\label{erfi}
\end{equation}
with
\begin{equation}
\mathcal C=\sqrt{\pi}\left(\frac{a\lambda_1}{\sqrt{s}}-(1-a)\sqrt{s}\right),
\end{equation}
where $\mathrm{erfi}$ is the imaginary error function. This expression is real for both signs of $s$. For $s<0$ it can be rewritten in terms of the error function. For $s\to0$ it reduces to Eq.~\eqref{RNlike}. Related exponential and error-function radial structures also occur in planar axionic Horndeski solutions~\cite{Baggioli:2021ejg}, although the matter action and the geometry of the transverse sections differ from those considered here.

At large $r$ we find
\begin{align}
-g_{tt}&\simeq1-a-\frac{2M}{r}+\frac{2a\lambda_1}{r^2},\nonumber\\
g^{rr}&\simeq1-a-\frac{2M}{r}+\frac{2\left[a\lambda_1-(1-a)s\right]}{r^2}.
\end{align}
The coefficients of $1/r^2$ in the two metric functions differ, so that the general branch is not of Reissner--Nordstr\"om type. This difference already appears in the asymptotic region, where the curvature is weak. To first order in the couplings, it is the difference between the metric $g$ and the metric $\hat g$ of Sec.~\ref{sec:redef}. For radial null vectors, Eq.~\eqref{Tcomp} gives $T_{\mu\nu}k^\mu k^\nu=-4s(k^r)^2/r^4$, independently of the sign of $F$. Therefore the radial null energy condition is violated for $s>0$, saturated for $s=0$ and satisfied for $s<0$. This is a property of the stress tensor defined with respect to $g_{\mu\nu}$. To first order in the couplings, the source in the metric $\hat g_{\mu\nu}$ of Sec.~\ref{sec:redef} saturates the radial null energy condition for both signs of $s$, in the same way as the energy conditions of scalar--tensor theories depend on the conformal frame. Near the center, $F(0)=a\lambda_1/s$ is finite for $s>0$, while the scalar curvature diverges as $R\simeq-8a\lambda_1s/r^6$. For $s<0$ the factor $e^{-s/r^2}$ diverges at the center, and $F(0)$ is finite only if the mass takes the value $2M=\sqrt{\pi}\left(a\lambda_1/\sqrt{|s|}+(1-a)\sqrt{|s|}\right)$. In this case we also find $F(0)=a\lambda_1/s$ and $R\simeq-8a\lambda_1s/r^6$. A finite $F(0)$ does not imply a regular center, and the solution is singular for both signs of $s$. For $\lambda_1=0$ the leading term vanishes, but we find $R\simeq4(1-a)s/r^4$ in the branches with finite $F(0)$, and the center remains singular. For $s<0$ and other values of the mass, $F$ diverges exponentially at the center. No $r^2$ term appears at large $r$ for constant couplings, so that the curvature couplings do not generate an effective cosmological constant.

The horizon structure can be obtained analytically. The condition $F(r_+)=0$ gives the mass as a function of the horizon radius,
\begin{equation}
M_h(r_+)=\frac{1}{2}\left[(1-a)r_+e^{s/r_+^2}+\mathcal C\,\mathrm{erfi}\!\left(\frac{\sqrt{s}}{r_+}\right)\right],
\end{equation}
and
\begin{equation}
\frac{dM_h}{dr_+}=\frac{1}{2}e^{s/r_+^2}\left(1-a-\frac{2a\lambda_1}{r_+^2}\right).
\label{dMdr}
\end{equation}
We take $0<a<1$. Extremal horizons occur at $r_e^2=2a\lambda_1/(1-a)$, which requires $\lambda_1>0$ and is the same as in the single-function branch. The extremal radius is independent of $s$, and therefore of $\lambda_2$, while the extremal mass $M_e=M_h(r_e)$ depends on both couplings. For $s>0$ and $\lambda_1>0$, $M_h$ diverges for $r_+\to0$ and $r_+\to\infty$, and has a minimum at $r_e$. There are two horizons for $M>M_e$, an extremal horizon for $M=M_e$ and none for $M<M_e$. For $\lambda_1\leq0$, $M_h$ increases monotonically from $-\infty$, and there is exactly one horizon. For $s<0$, $M_h$ tends at the center to the finite value
\begin{equation}
M_0=\frac{\sqrt{\pi}}{2}\left(\frac{a\lambda_1}{\sqrt{|s|}}+(1-a)\sqrt{|s|}\right),
\end{equation}
which is the mass that makes $F(0)$ finite. For $\lambda_1>0$ there is no horizon for $M<M_e$, an extremal horizon for $M=M_e$, two horizons for $M_e<M<M_0$ and one for $M\geq M_0$. For $\lambda_1\leq0$ there is no horizon for $M\leq M_0$ and one for $M>M_0$. Therefore, $M_0$ also marks a change in the number of horizons, from two to one for $\lambda_1>0$ and from none to one for $\lambda_1\leq0$. The extremal solutions are exact solutions of the action. Their interpretation as an approximation to a more fundamental theory, however, requires all the curvature corrections to be small at the horizon. As an estimate of scale, $\lambda_1/r_e^2=(1-a)/(2a)$, which is small only when $a$ is close to one.

The Hawking temperature is
\begin{equation}
T_H=\frac{N(r_+)F'(r_+)}{4\pi}=e^{s/r_+^2}\,\frac{(1-a)r_+^2-2a\lambda_1}{4\pi r_+^3},
\end{equation}
that is, the temperature of the single-function branch multiplied by $e^{s/r_+^2}$. The density of strings at the horizon is the same as before, and the Wald entropy is again $S=\pi\left[r_+^2-2a(\lambda_1+\lambda_2+\lambda_3)\right]$. From Eq.~\eqref{dMdr} we get $dM_h/dr_+=T_H\,dS/dr_+$, with $dS/dr_+=2\pi r_+$, so that the first law $dM=T_HdS$ holds analytically for both signs of $s$, at fixed $a$ and $\lambda_i$. As in the single-function branch, $M$ is the parameter in the metric, with $N\to1$ at infinity.

\section{Conclusions}\label{sec:conclusions}

In this work, we revisited the Letelier--Alencar cloud of strings and addressed the microscopic origin of its magnetic-like component. We showed that no ensemble of free Nambu--Goto strings with positive tensions and non-negative weights can produce a static spherically symmetric source with $p_r=-\rho$ and $p_\perp>0$. The argument uses only the null energy condition and does not assume that the strings are static, straight or non-intersecting. The non-simple bivector of the model is the dual of a field strength with $F\wedge F\neq0$, and the static source of the Letelier--Alencar cloud coincides with that of the dyonic sector of the square-root electrodynamics of Nielsen and Olesen, with $a$ the magnetic charge and $c_0^2$ the electric charge, in agreement with Refs.~\cite{Mazharimousavi:2012zx,Alencar:2026dsq}. Where $F\wedge F\neq0$, the equivalence also holds locally for the equations of motion. The analogy with DBI reproduces, in the strong-field regime, only the Letelier cloud with $p_\perp=0$. We also showed that, in the static spherically symmetric sector and in areal-radius gauge, any conserved source with $p_r=-\rho$ can be read as a cloud of radial strings whose tension depends on the density of strings. In this language, the tangential pressure measures the interaction between strings, and the Letelier--Alencar cloud corresponds to an interaction that starts at third order in the density. Within the string-fluid variational framework~\cite{Kopczynski:1987wb,Schubring:2014ena}, we obtained an effective action that reproduces the Letelier--Alencar stress tensor with integrable worldsheets and a simple flux two-form. Its energy density has a cubic leading correction to the free-string term, while its microscopic origin remains open. The same static source also admits an effective realization through a constrained scalar triplet with a noncanonical kinetic function, extending the nonlinear-sigma-model description of the global-monopole exterior.

We then derived transverse stresses from nonminimal couplings of the strings to the spacetime curvature in a worldsheet action. Each string couples to the spacetime curvature through its worldsheet, with the three couplings linear in the curvature that can be built with the metric and the projector on the worldsheet. Within the general framework for surface actions~\cite{Armas:2017pvj}, we derived the stress-energy tensor for arbitrary embeddings and showed that radial strings remain exact solutions of their equations of motion. The coupling to the Riemann tensor does not contribute to the source of radial strings, as follows from the Gauss identity, but it shifts the Wald entropy by a constant. For $\lambda_2=-3\lambda_1$ the metric has a single function, and for $\lambda_1>0$ the strings support a positive tangential pressure with a simple bivector. The geometry coincides with that of a charged black hole surrounded by a cloud of strings, but the $1/r^2$ term now comes from the strings themselves. In the general case the worldsheet action yields the two-function family, given in terms of the imaginary error function, which is the central geometric result of this work. To first order in the couplings, a metric redefinition maps the theory to general relativity with a contact interaction quadratic in the density of strings, and the two-function family is the corresponding geometry in the metric to which the strings couple. Its two metric functions are physically distinct when ordinary matter couples to this metric, and the exact form contains all orders in the couplings. Its horizon structure can be obtained analytically: the extremal radius is independent of $\lambda_2$, and for $s<0$ the mass that makes $F(0)$ finite also marks a change in the number of horizons. In both branches the first law is satisfied with the Wald entropy, and the center is singular. The nonminimal worldsheet action therefore provides an explicit self-gravitating string model with transverse stresses, consistent radial embeddings and an exactly integrable metric.

Several extensions are possible. A first direction concerns the dynamics of the effective realizations. The string-fluid and constrained scalar descriptions provide complementary realizations of the same static source, and their perturbations and stability are left for future work. Perturbations are also the natural place to distinguish the single-function branch from the charged black hole surrounded by a cloud of strings, since both share the same static geometry.

A second direction is to go beyond radial configurations. The tensor derived in Appendix~\ref{app:Tmunu} applies to deformed worldsheets and allows the study of circular and rotating strings. In this case contributions with different orientations at the same point must be considered, and the $\lambda_3$ coupling describes extrinsic curvature effects.

Finally, nonlinear curvature couplings, the regularity of the center and the conserved charges deserve further study. For radial worldsheets $\mathcal I_3={}^{(2)}\!R$, but $\sqrt{-\gamma}\,({}^{(2)}\!R)^2$ is not a total derivative, and such terms contribute to the radial field equations. Whether they admit regular solutions or reproduce exactly the Letelier--Alencar source remains to be investigated. Regular string-cloud geometries and dynamical-tension constructions have also been explored~\cite{Muniz:2025reg,Bahamonde:2026bnw}. A complete Hamiltonian treatment of the conserved charges on the conical background and the extended thermodynamics with varying $a$ and $\lambda_i$ are also interesting problems. In this work the couplings $\lambda_i$ are phenomenological parameters of the action, and relating them to strings of finite thickness would require a specific derivation.

In the equivalent fluid description of the single-function branch, the correction to the free-string energy density is quadratic in the string density. For the Letelier--Alencar source, the leading correction is cubic. This leaves a concrete microscopic question: which interactions between strings can produce the Letelier--Alencar equation of state?

\appendix

\section{Derivation of the Stress-Energy Tensor}\label{app:Tmunu}

In this appendix we derive the stress-energy tensor of the cloud for the three couplings in the action \eqref{action}, within the general framework for surface actions of Ref.~\cite{Armas:2017pvj}. Our conventions for the stress tensor and the curvature agree with those of Ref.~\cite{Armas:2017pvj}, whose tangential projector $\gamma^{\mu\nu}$ corresponds to our $P^{\mu\nu}$. We define $T^{\mu\nu}=(2/\sqrt{-g})\,\delta S_m/\delta g_{\mu\nu}$, where $S_m$ is the action of the strings, and write $h_{\mu\nu}=\delta g_{\mu\nu}$. The embeddings are kept fixed, and the variation is derived for each worldsheet separately. For the radial cloud considered in this work, all contributions at a given point share the same tangent projector, so that a single $P^{\mu\nu}$ appears in the source. For overlapping worldsheets with different tangent planes, the contributions must be summed with their respective projectors before any expression with a common projector is used. The matter action can be written as
\begin{equation}
S_m=-\int d^4x\sqrt{-g}\,\rho\left[1+\lambda_1R+\lambda_2\mathcal I_2+\lambda_3\mathcal I_3\right],
\end{equation}
where $\sqrt{-g}\,\rho=\mu\sum_i\int d^2\xi\sqrt{-\gamma_i}\,\delta^{(4)}(x-X_i)$ depends on the metric only through $\sqrt{-\gamma_i}$. From $\delta\gamma_{AB}=h_{\mu\nu}\partial_Ax^\mu\partial_Bx^\nu$ we get
\begin{align}
\delta\left(\sqrt{-g}\,\rho\right)&=\frac{1}{2}\sqrt{-g}\,\rho P^{\mu\nu}h_{\mu\nu},\nonumber\\
\delta P^{\mu\nu}&=-P^{\mu\alpha}P^{\nu\beta}h_{\alpha\beta}.
\label{deltarho}
\end{align}
For $\lambda_i=0$ the first relation gives $T^{\mu\nu}=-\rho P^{\mu\nu}$, the source of the Nambu--Goto cloud.

The terms with derivatives of $h_{\mu\nu}$ come from the variation of the curvature. The variation of the Christoffel symbols is
\begin{equation}
\delta\Gamma^\eta_{\alpha\beta}=\frac{1}{2}g^{\eta\sigma}\left(\nabla_\alpha h_{\sigma\beta}+\nabla_\beta h_{\sigma\alpha}-\nabla_\sigma h_{\alpha\beta}\right),
\label{deltaGamma}
\end{equation}
and the Palatini identity gives
\begin{equation}
\delta R^\eta_{\ \beta\gamma\delta}=\nabla_\gamma\delta\Gamma^\eta_{\delta\beta}-\nabla_\delta\delta\Gamma^\eta_{\gamma\beta}.
\label{PalatiniRiem}
\end{equation}
We use $[\nabla_\gamma,\nabla_\delta]v^\alpha=R^\alpha_{\ \beta\gamma\delta}v^\beta$, $R_{\alpha\beta\gamma\delta}=g_{\alpha\eta}R^\eta_{\ \beta\gamma\delta}$ and $R_{\beta\delta}=R^\alpha_{\ \beta\alpha\delta}$. In what follows we integrate by parts in spacetime and drop boundary terms, which vanish for variations $h_{\mu\nu}$ of compact support. No assumption is made about the embeddings, and the order of the covariant derivatives is kept at each step.

Consider a symmetric tensor $J^{\alpha\beta}$ and the integral $\int d^4x\sqrt{-g}\,J^{\alpha\beta}\delta R_{\alpha\beta}$. From Eq.~\eqref{PalatiniRiem} we have $\delta R_{\alpha\beta}=\nabla_\eta\delta\Gamma^\eta_{\alpha\beta}-\nabla_\beta\delta\Gamma^\eta_{\alpha\eta}$, and from Eq.~\eqref{deltaGamma} $\delta\Gamma^\eta_{\alpha\eta}=\frac{1}{2}\nabla_\alpha h$, with $h=g^{\mu\nu}h_{\mu\nu}$. A first integration by parts gives
\begin{align}
\int\! d^4x\sqrt{-g}\,J^{\alpha\beta}\delta R_{\alpha\beta}
=\int\! d^4x\sqrt{-g}\,\Big[&-\nabla_\eta J^{\alpha\beta}\,\delta\Gamma^\eta_{\alpha\beta}\nonumber\\
&+\frac{1}{2}\nabla_\beta J^{\alpha\beta}\,\nabla_\alpha h\Big].
\end{align}
Using Eq.~\eqref{deltaGamma} and the symmetry of $J^{\alpha\beta}$, the first term is $-\nabla^\sigma J^{\alpha\beta}\nabla_\alpha h_{\sigma\beta}+\frac{1}{2}\nabla^\sigma J^{\alpha\beta}\nabla_\sigma h_{\alpha\beta}$. A second integration by parts moves the remaining derivatives to $J^{\alpha\beta}$,
\begin{align}
&\int\! d^4x\sqrt{-g}\,J^{\alpha\beta}\delta R_{\alpha\beta}\nonumber\\
&=\int\! d^4x\sqrt{-g}\,\Big[\nabla_\alpha\nabla^\sigma J^{\alpha\beta}\,h_{\sigma\beta}\nonumber\\
&\quad-\frac{1}{2}\Box J^{\alpha\beta}\,h_{\alpha\beta}
-\frac{1}{2}\nabla_\alpha\nabla_\beta J^{\alpha\beta}\,h\Big].
\end{align}
Note that, in the first term, the derivative contracted with $J^{\alpha\beta}$ acts last. Symmetrizing in the free indices we obtain
\begin{equation}
\int\! d^4x\sqrt{-g}\,J^{\alpha\beta}\delta R_{\alpha\beta}=\!\int\! d^4x\sqrt{-g}\,\mathcal D^{\mu\nu}[J]h_{\mu\nu},
\label{Palatini}
\end{equation}
with $\mathcal D^{\mu\nu}[J]$ given in Sec.~\ref{sec:nonminimal}. Since the covariant derivatives do not commute, a different order would produce additional curvature terms.

Consider first the $\lambda_1$ term. Using $\delta R=-R^{\mu\nu}h_{\mu\nu}+g^{\alpha\beta}\delta R_{\alpha\beta}$ and Eq.~\eqref{Palatini} with $J^{\alpha\beta}=\rho g^{\alpha\beta}$, we find $\mathcal D^{\mu\nu}[\rho g]=\nabla^\mu\nabla^\nu\rho-g^{\mu\nu}\Box\rho$. Together with Eq.~\eqref{deltarho}, this gives
\begin{equation}
T^{\mu\nu}_{(1)}=\lambda_1\left[-\rho RP^{\mu\nu}+2\rho R^{\mu\nu}+2(g^{\mu\nu}\Box-\nabla^\mu\nabla^\nu)\rho\right].
\end{equation}

For the $\lambda_2$ term there are three contributions: the variation of the density, the variation of the projector and the variation of the Ricci tensor, with $J^{\alpha\beta}=\rho P^{\alpha\beta}$ in Eq.~\eqref{Palatini}. We obtain
\begin{equation}
T^{\mu\nu}_{(2)}=\lambda_2\left[2\rho R_{\alpha\beta}P^{\alpha(\mu}P^{\nu)\beta}-\rho\,\mathcal I_2P^{\mu\nu}-2\mathcal D^{\mu\nu}[J]\right].
\end{equation}
The sum of these two terms and the Nambu--Goto term gives Eq.~\eqref{Tcov}.

For the $\lambda_3$ term we write $\mathcal I_3=g_{\eta\alpha}R^\eta_{\ \beta\gamma\delta}P^{\alpha\gamma}P^{\beta\delta}$ and define
\begin{equation}
M^{\alpha\beta\gamma\delta}=\frac{1}{2}\left(P^{\alpha\gamma}P^{\beta\delta}-P^{\alpha\delta}P^{\beta\gamma}\right),
\end{equation}
which has the algebraic symmetries of the Riemann tensor and satisfies $R_{\alpha\beta\gamma\delta}M^{\alpha\beta\gamma\delta}=\mathcal I_3$. There are four contributions. The variation of the density gives $\frac{1}{2}\rho\,\mathcal I_3P^{\mu\nu}h_{\mu\nu}$. The variation of $g_{\eta\alpha}$ gives $\rho R^{\mu}_{\ \beta\gamma\delta}P^{\nu\gamma}P^{\beta\delta}h_{\mu\nu}$. The variation of the two projectors gives two equal terms, and their sum is $-2\rho R_{\alpha\beta\gamma\delta}P^{\alpha\mu}P^{\gamma\nu}P^{\beta\delta}h_{\mu\nu}$.

The last contribution comes from the variation of the Riemann tensor. Since $\delta R^\eta_{\ \beta\gamma\delta}$ is antisymmetric in $\gamma\delta$, only the antisymmetric part of $P^{\alpha\gamma}P^{\beta\delta}$ contributes, and from Eq.~\eqref{PalatiniRiem} we get
\begin{align}
\int\! d^4x\sqrt{-g}\,&\rho M_\eta^{\ \beta\gamma\delta}\delta R^\eta_{\ \beta\gamma\delta}\nonumber\\
&=2\!\int\! d^4x\sqrt{-g}\,\rho M_\eta^{\ \beta\gamma\delta}\nabla_\gamma\delta\Gamma^\eta_{\delta\beta}.
\end{align}
A first integration by parts gives $-2\int d^4x\sqrt{-g}\,V_\eta^{\ \beta\delta}\delta\Gamma^\eta_{\delta\beta}$, where
\begin{equation}
V^{\sigma\beta\delta}=\nabla_\gamma\left(\rho M^{\sigma\beta\gamma\delta}\right)
\end{equation}
is antisymmetric in $\sigma\beta$. Inserting Eq.~\eqref{deltaGamma}, the term with $\nabla_\delta h_{\sigma\beta}$ vanishes by this antisymmetry. Exchanging $\sigma$ and $\beta$ in the term with $\nabla_\sigma h_{\delta\beta}$, it becomes equal to the remaining one, and we get $-2\int d^4x\sqrt{-g}\,V^{\sigma\beta\delta}\nabla_\beta h_{\sigma\delta}$. A second integration by parts gives
\begin{equation}
2\!\int\! d^4x\sqrt{-g}\,\nabla_\beta\nabla_\gamma\left(\rho M^{\mu\beta\gamma\nu}\right)h_{\mu\nu},
\end{equation}
where the order of the derivatives is the one obtained in the two steps. Collecting the four contributions, we find
\begin{align}
T^{\mu\nu}_{(3)}=\lambda_3\Big[&-\rho\,\mathcal I_3P^{\mu\nu}-2\rho R^{(\mu}_{\ \ \beta\gamma\delta}P^{\nu)\gamma}P^{\beta\delta}\nonumber\\
&+4\rho R_{\alpha\beta\gamma\delta}P^{\alpha\mu}P^{\gamma\nu}P^{\beta\delta}\nonumber\\
&-4\nabla_\beta\nabla_\gamma\left(\rho M^{(\mu|\beta\gamma|\nu)}\right)\Big].
\label{T3}
\end{align}
This expression is valid for arbitrary embeddings. For totally geodesic worldsheets it must vanish in the bulk, up to boundary contributions, as discussed in Sec.~\ref{sec:nonminimal}. For deformed worldsheets it gives the extrinsic curvature contributions of the $\lambda_3$ coupling.

\acknowledgments
The authors thank the Conselho Nacional de Desenvolvimento Cient\'{i}fico e Tecnol\'{o}gico (CNPq) and the Funda\c{c}\~ao Cearense de Apoio ao Desenvolvimento Cient\'{i}fico e Tecnol\'{o}gico (FUNCAP).

\bibliographystyle{JHEP}
\bibliography{refArxiv_rai}
\end{document}